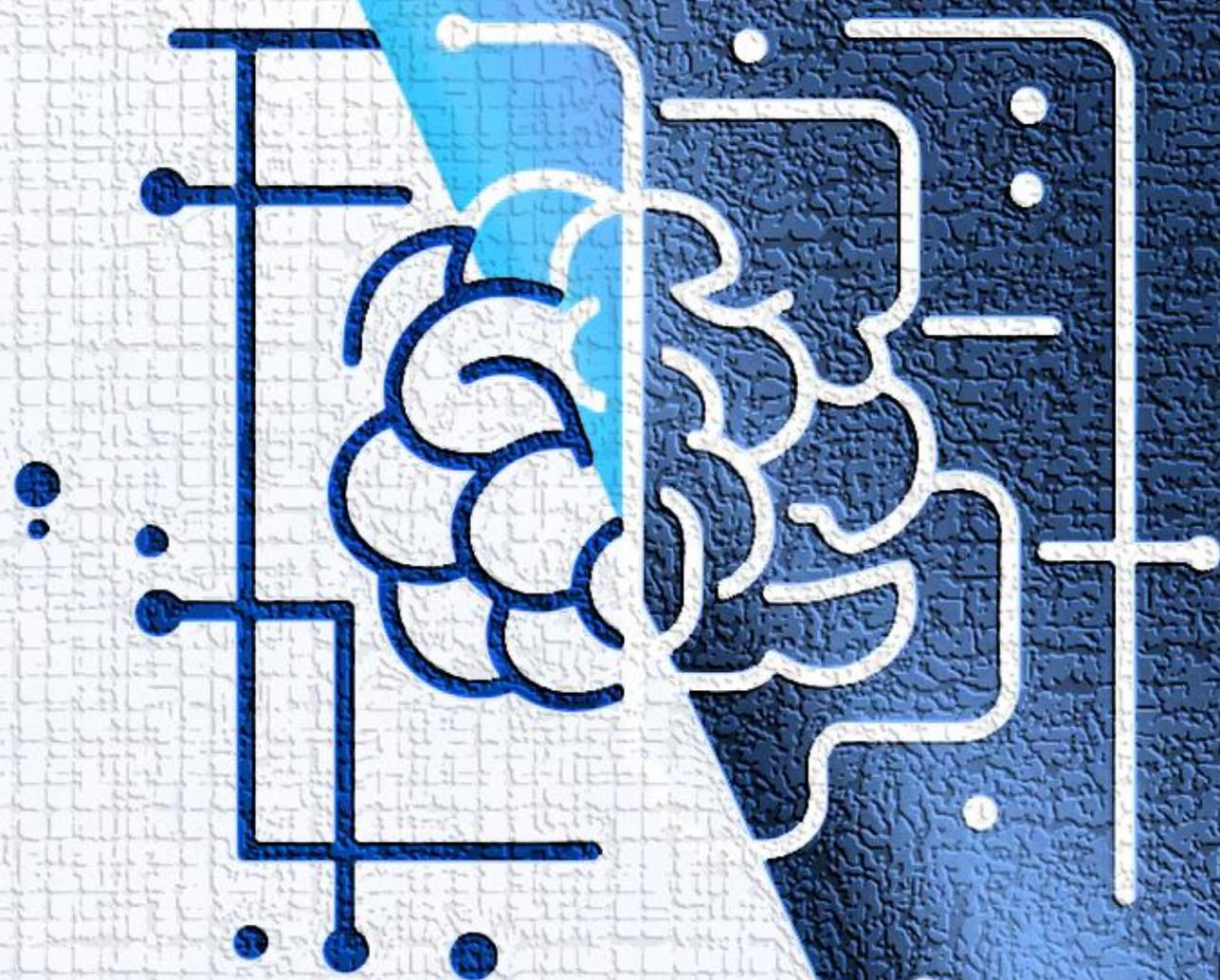

# Literacies and Labor Market Challenges in the Digital Era

**Chapter 11. Mapping Literacies in the Tourism Labor Market: A Cross-Database Comparison**

Eddy Soria Leyva[a], Aïda Valls Mateu[a], Ana Beatriz Hernández Lara[a*]

[a] Universitat Rovira i Virgili, Spain
[*] anabeatriz.hernandez@urv.cat

**Abstract**

This book chapter conducts a comparative bibliometric analysis of literacies in the tourism labor market, drawing from the Web of Science (WoS) and Scopus databases. Its aim is to assess scientific output and identify key patterns of scientific production and collaboration. Findings show contrasts between the coverage of the two databases and their scientific output, with an overlap of 35.71%. A gradual and correlated increase in the number of publications over time is observed. Scopus stands out for its broader impact and enduring citation relevance, suggesting its academic contributions have a longer-lasting effect. Conversely, WoS is characterized by a focus on more recent influential publications and exhibits a marginally more intense collaboration network.

# 1. Introduction

The tourism sector stands as a cornerstone of the global economy (Alsahafi et al., 2023; Buhalis et al., 2023; de Pablo Valenciano et al., 2023), representing 10% of global gross domestic product (GDP) and having a long track record of contributing to global sustainable development (Buhalis et al., 2023). It emerges as one of the fastest-growing sectors (Jeong et al., 2023; Mihigo & Lukenangula, 2023), significantly stimulating economic growth through income generation, investments, and exports, and playing a crucial role in creating jobs across diverse geographical regions. In particular, the labor market in tourism is characterized by its diversity, including but not limited to activities in hospitality, travel services, leisure and recreation, and cultural and natural heritage sites, while providing a wide range of opportunities that span various skill levels and professions.

The demand for a skilled workforce in the tourism sector is driven by the need to meet evolving consumer expectations, technological advancements, and the imperative for sustainable tourism practices (Chen & Yu, 2023; Gössling, 2021; Sharpley, 2021). In this context, the concept of literacies becomes critically important. Literacies extend beyond the traditional understanding of reading and writing skills to encompass a broad set of competencies, knowledge areas, and capabilities (Benavides Rincón & Díaz-Domínguez, 2022; Cicha et al., 2021; Long & Magerko, 2020; Pothier & Condon, 2020; Wuyckens et al., 2022). In the tourism sector, literacies encapsulate not only the basic skills of communication and numeracy, including also digital literacy, cultural and environmental sensitivity, language proficiency, and interpersonal competencies (Apelt et al., 2023; Caldevilla-Domínguez et al., 2021; Gössling, 2021; Johnson, 2014; Makandwa et al., 2023). Literacies in the tourism sector are essential given its complexity, its technological innovations and digitalization, the engagement that entails with diverse cultural backgrounds, and the need to contribute to sustainable tourism practices (Jeong et al., 2023; Sharpley, 2021).

However, while several bibliometric studies have explored specific types of literacies, there is a critical gap in our understanding of the overall research landscape and intellectual structure of the field of literacies within the tourism labor market. The existing studies often focus on niches such as Information and Communication Technology (Caldevilla-Domínguez et al., 2021), Tourism Education Advancement (Au-Yong-Oliveira et al., 2024; Yağmur Karalım, 2024), Digital Literacy (Caldevilla-Domínguez et al., 2021; Dinis et al., 2022), Digital Transformation (Bekele

& Raj, 2024; Indahyani et al., 2023; S. Kumar et al., 2023), Literacy Scales (Koç et al., 2023; Lin et al., 2023; Luna-Cortes, 2024; Luna-Cortés & Brady, 2024), Animal Welfare Literacy (Fennell, 2022) and e-Tourism Information Literacy (Wang et al., 2023). However, bibliometric studies on literacies in the general context of the tourism labor market are still in their early stages of development.

In fact, Luna-Cortes (2024) recently noted that the differences between previous studies associated with literacy constructs in tourism lead to a need for a structured overview of prior research. Also, practitioners need to stay informed about the latest research findings and trends to ensure that their training efforts are in line with the most relevant developments in the industry. Therefore, we conducted a bibliometric study on publications indexed in the Web of Science (WoS) and Scopus, as without a thorough understanding of core literacy trends in the tourism labor market, training programs risk being misaligned, tourists underserved, and sustainable practices compromised.

Comparing different databases is critical for researchers as it ensures identifying unique and overlapping resources (A. Kumar, 2021; Pranckutė, 2021; Sánchez et al., 2017). Consequently, this cross-database bibliometric study addresses the aforementioned shortcomings by providing a holistic view of the current state of scientific research on literacy in the tourism sector, revealing both basic patterns and trends in research. Thus, this study contributes to a deeper understanding of the intellectual structure of the field, identifies key sources and authors, and provides insights into collaboration, topic and citation patterns. To the best of our knowledge, this is one of the first studies to perform a cross-database comparison of literacies within the general context of the tourism labor market.

Our research focuses on two fundamental issues, which are expressed through the following research questions:

> RQ1: How do performance metrics and key academic contributions vary in the field of literacies within the tourism sector between the WoS and Scopus databases?
>
> RQ2: What patterns of scientific collaboration in the field of literacies within the tourism sector emerge from a joint analysis of the Scopus and WoS databases?

## 2. Literature Review

### 2.1. Literacies and tourism labor market

Traditionally, literacy was understood as the ability to read and write. However, the concept has evolved in the current digital age, as the emergence of the knowledge-based society implies that every citizen must possess basic skills and be 'digitally literate' in order to have an equal footing in terms of job opportunities (Ng et al., 2021). Dobryakova et al. (2023) analyzed over 180 national and international frameworks of competences and "twenty-first century skills", including specialized reports by the European Commission, OECD, UNESCO, World Economic Forum, ATC21S, P21, EnGauge and others, as well as comparative reviews, to arrive at a definition of literacies as abilities in specific domains, applying skills in meaningful contexts.

The concept has now been extended to new literacies such as environmental (Afandi et al., 2023; Zheng et al., 2020), STEM (Schirone, 2022), financial (Fengwen & Ali, 2023; Gallego-Losada et al., 2022; Goyal & Kumar, 2021), physical (Kokot & Turnšek, 2022; Mendoza-Muñoz et al., 2022), health (Sun et al., 2021; Wilson et al., 2022), media (Bapte, 2021; Kutlu-Abu & Arslan, 2023; Park et al., 2021), digital (Caldevilla-Domínguez et al., 2021; Park et al., 2021; Purnomo et al., 2020; Sharma et al., 2023; Sujarwo et al., 2022), information (Bapte, 2020; Park et al., 2021), and AI literacy (Ng et al., 2021; Tenório et al., 2023).

After reviewing the new literacies that have emerged, we propose a notion of literacy that can be better defined as a comprehensive construct that not only includes the fundamental skills of reading and writing but also embodies the critical and analytical abilities to comprehend, interpret, and interact with diverse content across different socio-cultural and digital media landscapes, encapsulating a range of competencies from simple word decoding to complex information management, creation, synthesis, and communication. However, it is difficult to establish a clear taxonomy, as many new literacies contain overlapping areas and have been branching out greatly in recent years.

Particularly research on literacies in the tourism sector has been addressed through various thematic lenses, as evidenced by the multidisciplinary nature of the studies, including human resource management (Costa et al., 2020; Lee-Ross & Pryce, 2010; Yu & Liu, 2019), leadership competencies (Alam et al., 2023; Idris et al., 2022; Testa & Sipe, 2012), and the alignment between educational outcomes and industry requirements (Beh & Shah, 2017; Gjedia & Ndou, 2019). Its evolution reflects changes in industry demands, technological advancements, and socio-economic trends, highlighting a transition from traditional competencies to a broader understanding, incorporating digital literacy (Vasconcelos & Balula, 2019), sustainability (Carlisle et al., 2023; Davidson & Wang, 2011), employee well-being (Miyakawa & Oguchi, 2022), etc.

Literacies hold significant importance for tourism industry professionals, empowering them to adeptly occupy roles that demand a diverse array of skills, ranging from linguistic fluency (language literacy) to competencies in navigating modern technology. For instance, rapidly evolving digital technologies within tourism (Dutta & Das, 2023; Kim, 2023) highlight the need for an adaptable workforce skilled in digital tools and platforms. This transformation showcases the industry's responsiveness to technological innovation, requiring employees to effectively manage and engage with developments like service robots (Herawan et al., 2023; Kim, 2023) and digital management systems (Madzík et al., 2023). Recent research also provides evidence of the importance of the literacies required to work in the tourism sector amidst the growth and expansion of sustainable tourism products (Brannon, 2023; Buhalis et al., 2023; Carlisle et al., 2023), or the current need to enhance well-being and health (Font-Barnet & Nel-lo Andreu, 2023; Miyakawa & Oguchi, 2022), especially in the post-covid scenario (Quiroz-Fabra et al., 2023). Consequently, literacies are at the backbone of professional development in today's tourism labor market.

**2.2. Bibliometric analysis overview**

Bibliometrics is a quantifiable informatics strategy that evaluates developing patterns and the knowledge structure within a particular sector to collect quantifiable, reproducible, and objective data (Alhashmi et al., 2024). Bibliometric approaches are increasingly central to literature reviews in many disciplines, have gained immense popularity in business research in recent years (Donthu et al., 2021), and are widely used to analyze and visualize the state, structure, hotspots and trends of research in specific fields (Ganjihal et al., 2023; Veiga-del-Baño et al., 2023).

Bibliometric analysis techniques fall into two primary categories: (1) performance analysis, and (2) science mapping. In essence, performance analysis accounts for the contributions of research constituents, whereas science mapping focuses on the relationships between research constituents (Donthu et al., 2021). Some examples of performance indicators include publication related metrics (total publications, productivity per year, etc.), citation-related metrics (total citations, average citations, etc.), and citation-and-publication related metrics such as the collaboration index, h-index, g-index (Donthu et al., 2021), AWCR-index (del Barrio-García et al., 2020), and the i10-index (Donthu et al., 2021). While the techniques for science mapping include citation analysis, co-citation analysis, bibliographic coupling, co-word analysis, and co-authorship analysis (Donthu et al., 2021). Currently, various databases and academic search engines facilitate the search and retrieval of academic records, such as Scopus, WoS, Google Scholar, Dimensions, and Microsoft Academic (Martín-Martín et al., 2021), while others specialize in providing information on the impact and dissemination of research in digital media and social networks (altmetrics), such as Altmetric and Plum Analytics.

## 3. Methodology

### 3.1 Data collection and inclusion-exclusion criteria

This research used Scopus and WoS, as they are the two bibliographic databases generally accepted as the most comprehensive sources for various academic purposes (Pranckutė, 2021). A standardized search strategy was implemented to replicate it across both databases, ensuring that the results were matched. Initially, several preliminary searches were conducted to refine the query more precisely until relevant results for the study were obtained. The final query was: ("literac*" OR "competenc*" OR "skill*") AND ("touris*" OR "hotel*" OR "travel*" OR "service*" OR "hospitality") AND ("occupa*" OR "employ*" OR "job*" OR "work" OR "wage" OR "Talent Market" OR "Human Capital Market" OR "Manpower Market" OR "Labor Market").

The exported ".bib" files were merged, eliminating duplicate records, to create a combined result (Scopus+WoS) named "Merged Dataset". To remove duplicates from Scopus and WoS, the "duplicatedMatching()" function from the R Bibliometrix library was used (Aria & Cuccurullo, 2017). In this way, the .bib files from both Scopus and WoS were converted to "Data Frames", thus normalizing the bibliographic metadata tags. The restricted Damerau-Levenshtein distance was used to find duplicated documents, whose heuristic has been successfully proven in previous studies (Carè & Weber, 2023; Martín-Martín et al., 2021).

The terms "service*" and "hospitality" were included even though not all the publications retrieved were related to tourism, but once relevant publications were identified, it was preferred to add these keywords and then process the dataset through a two-step screening process (**Figure 1**), which allowed for refinement of the results. The scope was restricted to titles to exclude non-relevant results detected through preliminary searches. In addition, some areas of science not relevant to the study were excluded (Example: Medicine, Nursing, Biochemistry, Genetics and Molecular Biology, Physics and Astronomy, etc.). The search included all available time periods, as well as all types of documents.

**Figure 1**

*Search strategy for identifying relevant papers (PRISMA flow diagram)*

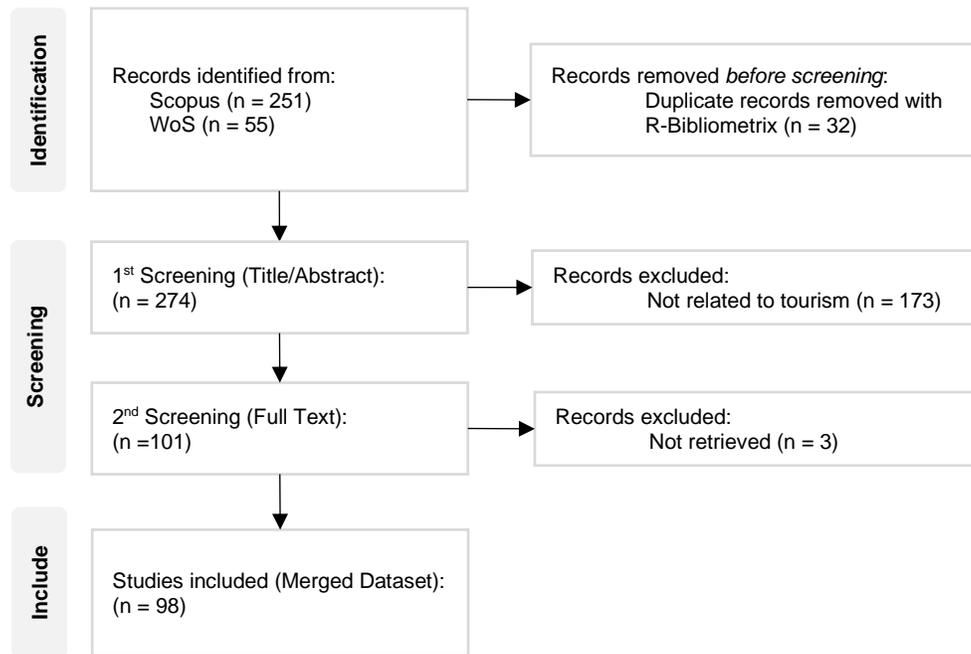

**3.2 Measurement and General Performance Analysis**

With the aim of answering research question RQ1, bibliometric indicators were considered to evaluate the performance of scientific production through the Scopus and WoS databases (number of authors, sources, geographical distribution of the number of publications, total number of citations, the h-index, g-index, i10-index, and the AWCR-index). The analysis was conducted using Python, a versatile and powerful programming language widely used for data analysis and visualization. Python scripts were developed and executed in Visual Studio Code, a free and open-source editor. The data frame generated with R-Bibliometrix was used as input and its preprocessing, filtering and manipulation were performed with the Python Pandas library. For example, the h-index was calculated by sorting the data based on citations and applying the threshold criteria that at least "h" number of papers have "h" or more citations. Some graphical representations were created with the Python libraries Matplotlib and Seaborn. For instance, the growth of scientific production was visualized using a simple line graph that plotted the number of publications per year.

**Table 1**

*Bibliometric indicators in scientific information platforms*

| Metrics | Definition | Scopus | WoS | Google Scholar | Plum Analytics |
|---|---|---|---|---|---|
| **SJR** | The SCImago Journal Rank (SJR) measures the prestige of citations received by a journal. | x | | | |
| **FWCI** | The Field-Weighted Citation Impact measures how an entity's publication citations compare to the average in its field. | x | | | |
| **SNIP** | The Source Normalized Impact per Paper quantifies citations received against expected citations in a serial's subject field. | x | | | |
| **CiteScore** | CiteScore measures average citations received per document published in the serial. | x | | | |
| **% First Author** | Percentage of publications where a researcher is the first author. | x | | | |
| **Quartile** | It ranks journals into four categories (Q1, Q2, Q3, Q4) based on their citation metrics, with Q1 representing the top 25%. | x | x | | |
| **NES** | The Normalized Eigenfactor Score (NES) measures journal influence through citation network density over a five-year period. | | x | | |
| **JIF** | The Journal Impact Factor (JIF) quantifies the average number of citations received by articles published in a particular journal over a specific period. | | x | | |
| **Edition** | Specific databases within WoS, including Science Citation Index Expanded (SCIE), Social Sciences Citation Index (SSCI), Arts & Humanities Citation Index (AHCI), Emerging Sources Citation Index (ESCI) and Conference Proceedings Citation Index (CPCI) | | x | | |
| **5-IF** | The 5 Year Impact Factor is calculated by dividing the number of citations in the JCR (Journal Citation Reports) year by the total number of articles published in the five previous years. | | x | | |
| **JCI** | The Journal Citation Indicator (JCI) is the average Category Normalized Citation Impact of citable items published by a journal over a recent three-year period. | | x | | |
| **5-Year Citations** | Number of new citations in the last 5 years. | | | x | |
| **h-index** | Largest number h such that h publications have at least h citations. | x | x | x | |
| **h5-index** | H-index for articles published in the last 5 complete years. | | | x | |
| **h5-median** | Median number of citations for the articles that make up its h5-index. | | | x | |
| **i10-index** | Number of publications with at least 10 citations. | | | x | |
| **Normalized TC** | This metric adjusts citation counts of a researcher's works based on their publication age. | x | | | |
| **g-index** | Highest number "g" such that the top "g" publications have, in total, at least "$g^2$" citations. | x | x | x | |
| **m-index** | H-index divided by the number of years the researcher has been active. | x | x | x | |
| **e-index** | A complement to the h-index, focusing on the excess citations beyond those covered by the h-index. | x | x | x | |
| **AWCR-index** | The Author Weighted Citation Rate (AWCR) is a metric reflecting the average citation rate per paper, weighted by the author's contribution to each paper. | x | x | x | |
| **PlumX Metrics** | Views (usage measure), Captures (bookmarks, saves), Social (social media likes and shares). | | | | x |

An overview of the performance indicators we used in this research is given in **Table 1**. It should be noted that both Google Scholar and Plum Analytics were incorporated into the comparison with the purpose of providing an external measure that enriches the analysis, both from the perspective of academic impact and social media presence.

**3.3 Analysis of Core Academic Contributions**

The core sources were identified by using Bradfords' Law, that states: "if journals in a field are sorted by number of articles into three groups, each with about one-third of all articles, then the number of journals in each group will be proportional to 1:n:n2" (Orăștean & Mărginean, 2023). The sources were divided into three zones according to Bradford's Law: Zone 1 (Core Sources), Zone 2 (Middle) and Zone 3 (Minor). The first nucleus zone contains a small number of highly productive journals say n1, the second zone contains a larger number of moderately productive journals, say n2, and the third zone containing a still larger number of journals of low productivity, say n3 (Ganjihal et al., 2023).

In this manner, these core sources were evaluated based on Google Scholar's metrics (h5-Index, h5-Median), Scopus' metrics (CiteScore, h-index, SJR, SNIP, FWCI, Quartile, Rank), and WoS' metrics (Edition, Quartile, JIF, JCI, NES and 5-IF). Similarly, the most cited papers and the most relevant authors were assessed, by using performance bibliometric indicators such as FWCI, Total Citations (TC), TC per Year, 5-year Citations, Normalized TC, Journal Quartile, PlumX metrics (Captures, Social, Views), h-index and i10-index.

**3.4 Co-authorship Network Analysis and Thematic Analysis**

We utilized VOSviewer's visualization capabilities (Ji et al., 2023) to conduct a co-authorship network analysis (to provide insights on research question RQ2). We used VOSviewer's Overlay Visualization and Density Visualization, defining association strength as the normalization method. In the Layout we used the default values provided by Vosviewer. To improve the visualization, we used the citations as weights, so that not only the collaborations but also the scientific impact could be clearly observed. The co-authorship network provides an overview of research collaborations within a field, as well as a visual representation of detached regions of research (Cardiff & Demirdžić, 2021; Higaki et al., 2020). Finally, to explain the patterns found through co-authorship analysis, we rely on the examination of thematic maps. The thematic map is based on density and centrality, with four predefined clusters that were generated with the R-bibliometrix software following Kaur et al (2024).

## 4. Findings and Discussion

### 4.1 Performance Analysis

#### *4.1.1 Performance Metrics*

A comparative analysis was conducted to address insights on research question RQ1, specifically in relation to the first part of the question, which focuses on describing performance metrics. Within the timespan from 1979 to 2023, a total of 98 unique papers were identified (Merged Dataset) as mentioned in the Method section. Scopus has indexed 78 articles while WoS has indexed 55, with 35 articles common to both (**Figure 2**).

**Figure 2**

*Venn Diagram representation of Merged Dataset (n=98)*

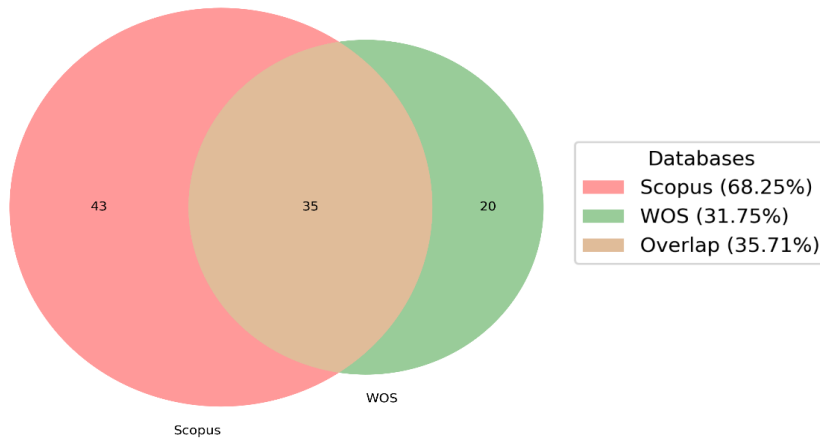

The annual growth rate of papers in the Merged Dataset is 6.35%, which is modest compared to the higher growth rate observed in WoS alone (10.53%). Conversely, papers from the Merged Dataset tend to be cited more frequently than those from WoS, with an average of 21.21 citations per document, suggesting a significant relevance of research in this field, an increase attributable to the Scopus database which has a broader coverage of sources.

In terms of document types, articles are the most common (74%), while the presence of conference papers (18%), book chapters (5%) and other types (3%) suggests a dynamic discourse, where findings are shared in various academic settings. The timespan of papers from the Merged

Dataset is significantly larger (45 years) compared to WoS (13 years), indicating a greater interest in this field in the sources indexed in Scopus. Overall, the results reflect a field that is not only growing as evidenced by citation rates, but also that the research has contrasts between databases (**Table 2**).

**Table 2**

*Comparative performance metrics*

| Metrics | Scopus | WoS | Merged Dataset |
|---|---|---|---|
| Timespan | 1979-2023 | 2000-2023 | 1979-2023 |
| Authors | 184 | 125 | 235 |
| Authors of single-authored docs | 17 | 13 | 23 |
| Sources | 59 | 46 | 77 |
| References | 4,364 | 2,438 | 4,998 |
| Citations | 2,053 | 419 | 2,079 |
| Years | 45 | 13 | 45 |
| Cites per Year | 45.62 | 32.23 | 46.20 |
| Authors per Paper | 2.36 | 2.27 | 2.40 |
| h-index (Hirsch's) | 23 | 11 | 23 |
| g-index (Egghe's) | 44 | 20 | 44 |
| i10-index | 34 | 12 | 35 |
| m-index | 0.51 | 0.85 | 0.51 |
| e-index | 35.01 | 14.93 | 35.01 |
| AWCR-index | 3.28 | 1.84 | 2.74 |

According to **Table 2**, the Merged Dataset encompasses 77 sources (after combining Scopus (59) and WoS (46) without duplicates) and Scopus's larger number of authors (184) compared to WoS (125) reflects a broader collaborative network. Besides, Scopus records a substantial citation count (2,053), which markedly exceeds that of WoS (419). This margin could emanate from differences in database coverage and citation practices, although it was possible to identify the high presence of some seminal works within the Scopus dataset.

Consequently, the h-index and g-index values from Scopus (h=23, g=44) are twice as high as those from WoS (h=11, g=20) and the Merged Dataset shows 35 papers with at least ten citations each (i10-index=35), similarly to Scopus (i10-index=34).

Lastly, the AWCR index considers both the quantity and longevity of citations, with Scopus showing the highest index (3.28). This implies that papers within the Scopus dataset not only receive numerous citations but also that these citations persist over time, indicating a lasting scholarly influence. However, the m-index, which adjusts the h-index for the career length of a researcher, is notably higher in WoS (0.85) than in the Scopus dataset. This could imply a more vibrant and active research output within WoS's timeframe or a concentration of influential work in more recent years.

**Figure 3** explores academic publication counts over the years. Prior to 2004, the presence of literature on the topic was minimal, indicating that this subject was underexplored. Subsequent years marked the beginning of a more consistent presence of research on this topic in both databases, with Scopus indexing more publications than WoS. This could be ascribed to Scopus's more extensive coverage in both this specific field and more broadly, potentially leading to a certain inherent bias regardless of the preference to publish in either database. Overall, the data reveal a gradual increase in the number of publications over the years, with both databases showing upward trends.

**Figure 3**

*Growth of scientific production*

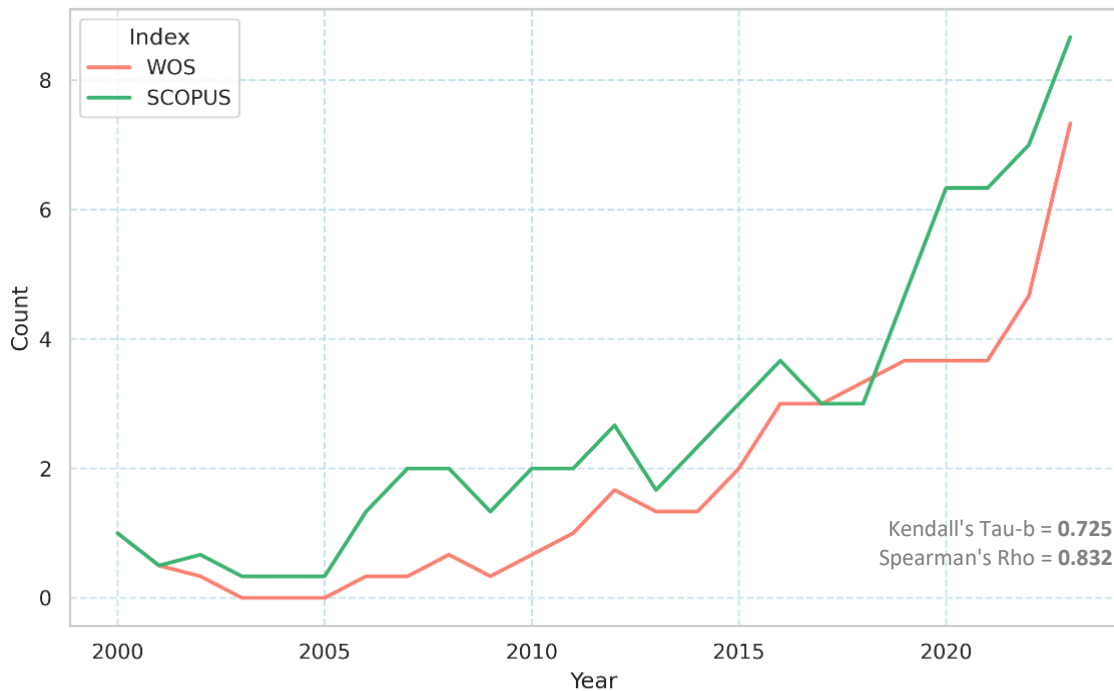

**Figure 4** shows the geographical distribution of these publications. In the Scopus dataset, Asia exhibits a robust network of scholarly production, with China at the forefront. European and North American regions, particularly the UK and the USA, also demonstrate high frequencies. Notably,

African, and Middle Eastern countries, such as South Africa and Egypt, indicate emerging participation in global scientific endeavors. Conversely, the WoS dataset reveals Ukraine sharing the lead with China, a striking difference from the Scopus data where Ukraine was absent. This difference is explained by the fact that the coverage among databases is different, and in this case, there are several publications made by Ukrainian researchers that were published in journals not indexed in Scopus, but in the WoS Core Collection (WoS-ESCI), such as "*Revista Românească pentru Educaţie Multidimensională*" and "*Journal for Educators, Teachers, and Trainers*". The frequency of publications from Turkey, Malaysia, and South Korea is consistent across both databases. Interestingly, the UK shows a diminished presence in WoS, and the USA appears significantly less when compared to the Scopus data. The reasons for these discrepancies might also be explained by the coverage of the databases, which is higher in Scopus.

**Figure 4**

*Geographic distribution of scientific production*

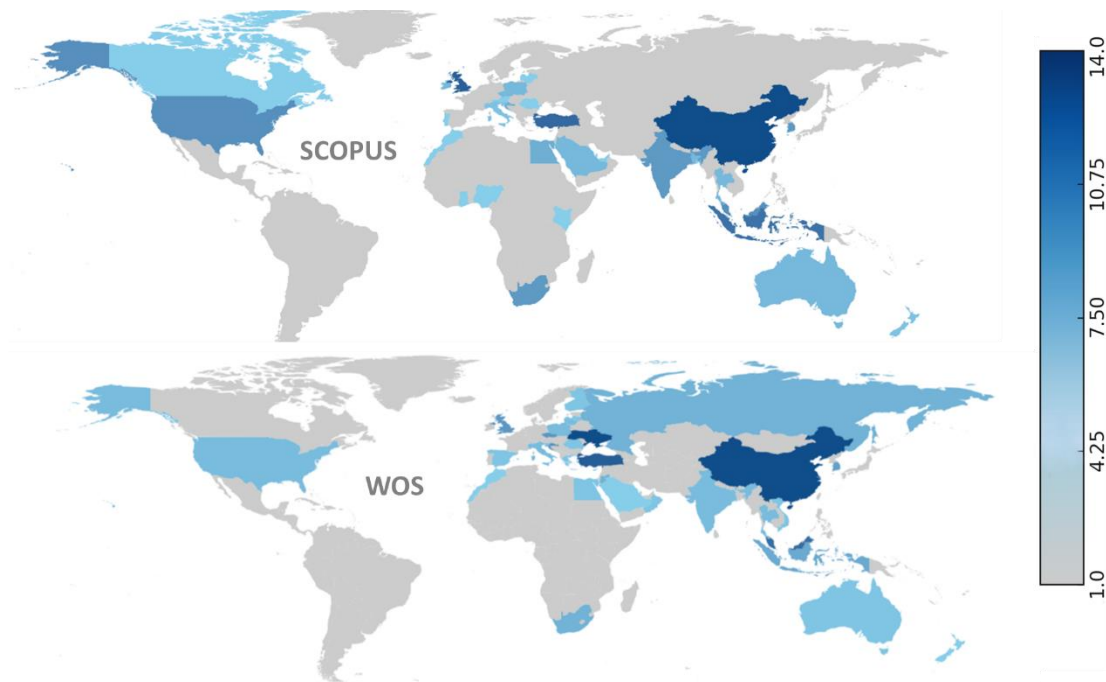

**Figure 5**

*Average citations per year*

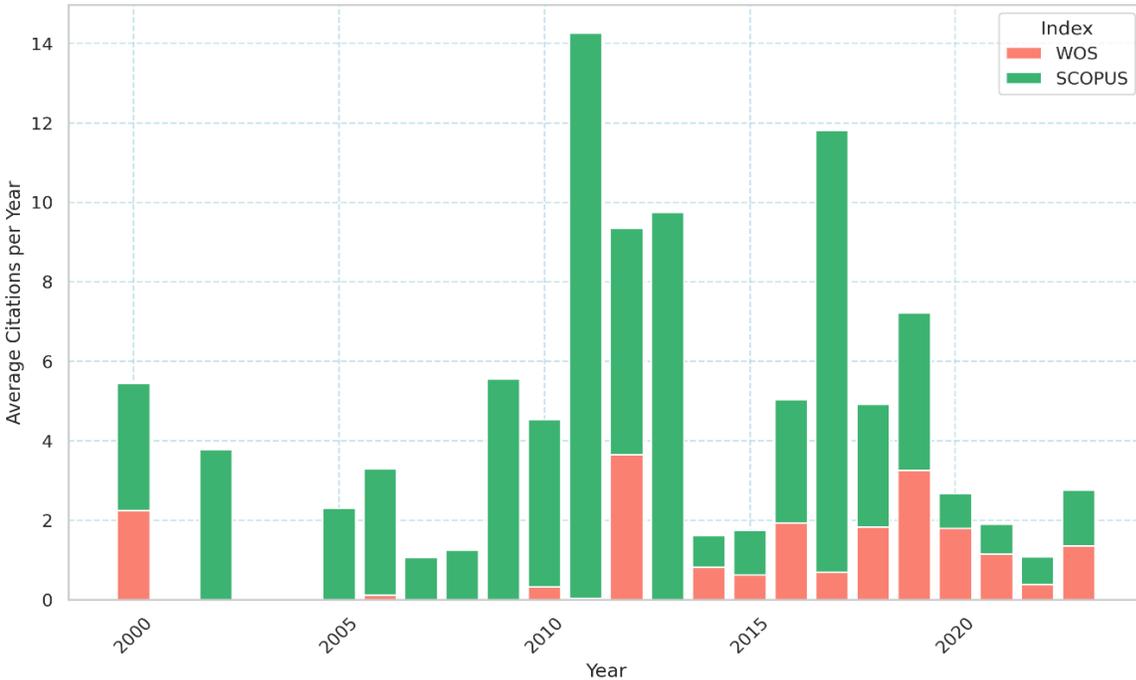

On the other hand, another performance metric to analyze was the distribution of the number of citations across databases. According to **Figure 5**, significant peaks in citation counts were observed: Scopus highlighted in the years 2011 and 2017, whereas WoS showed a similar pattern in 2012 and 2019. These peaks correspond to the publication of research with substantial impact, as evidenced by works from Chang et al. (2011), Daniel et al. (2017), Kong et al. (2012), and Safavi & Bouzari (2019). As can be seen, Scopus shows higher average citations during the beginning of the series, but in the last 5 years, it is WoS that has a higher weight, confirming the results suggested by the m-index.

### 4.2 Core Academic Contributions

The development of this section seeks to answer to research question RQ1, specifically in relation to key scholarly contributions, so that Core Sources, Key Papers and Top Authors will be addressed.

*4.2.1 Core Sources*

Bradford's Law posits that scientific papers are distributed in a manner where a few journals publish the majority of significant papers in a field, and as one moves away from this core, more journals publish fewer relevant papers. In **Figure 6**, the journals classified in Zone 1 according to Bradford's Law are shown.

Within the core group, the following 13 journals were identified: International Journal of Hospitality Management (IJHM), International Journal of Contemporary Hospitality Management (IJCHM), Anatolia (ANAT), Journal of Hospitality and Tourism Education (JHTE), African Journal of Hospitality, Tourism and Leisure (AJHTL), Investigaciones Turísticas (INVTUR), Journal of Environmental Management and Tourism (JEMT), Journal of European Industrial Training (JEIT), Journal of Hospitality and Tourism Insights (JHTI), Journal of Human Resources in Hospitality and Tourism (JHRHT), Journal of Teaching in Travel and Tourism (JTTT), Revista Romaneasca Pentru Educatie Multidimensionala (RRPEM) and Tourism Management (TOURMAN).

**Figure 6**

*Core Sources by Bradford's Law*

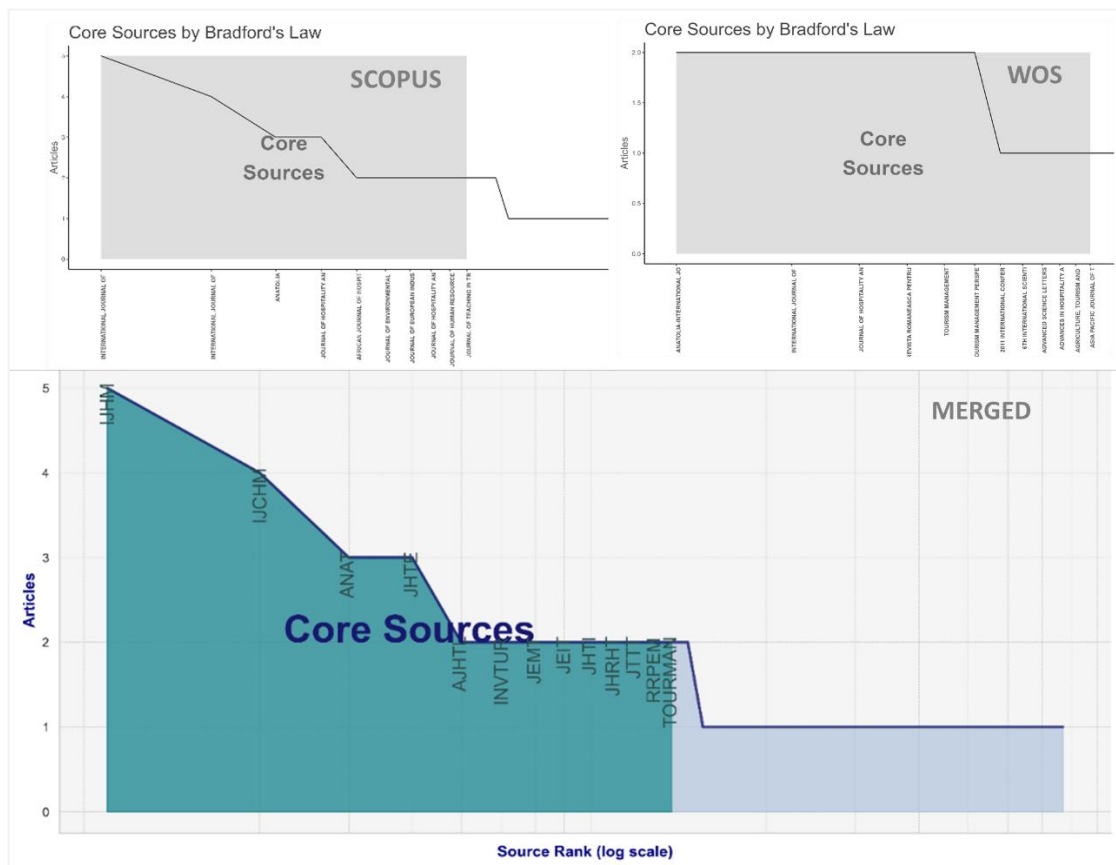

**Figure 7** shows the cumulative number of publications from the core sources. There is a general trend of growth, however, the contributions have fluctuated over the years. Some journals, represented by the bottom segments, have a longer history of publications such as IJCHM, TOURMAN and JEIT, while others have emerged more recently in this field: AJHTL, JEMT and RRPEM. IJHM and IJCHM stand out as the top publishers with 5 and 4 articles respectively, while INVTUR and JHTI have demonstrated the most rapid growth over the past 3 years in this research domain. **Table 3** presents a summary of the key metrics for all of these top core sources, offering a comparison between Scopus and WoS, and enhancing the analysis with metrics from Google Scholar.

According to **Table 3**, four categories were identified in which the core sources are distributed: Tourism (61%), Education (23%), Geography (8%), and Management (8%). TOURMAN and IJHM are the leading core journals based on their high CiteScores, H-Indices, Impact Factors and rankings across Scopus, WoS and Google Scholar.

**Figure 7**

*Cumulative Core Source Dynamics*

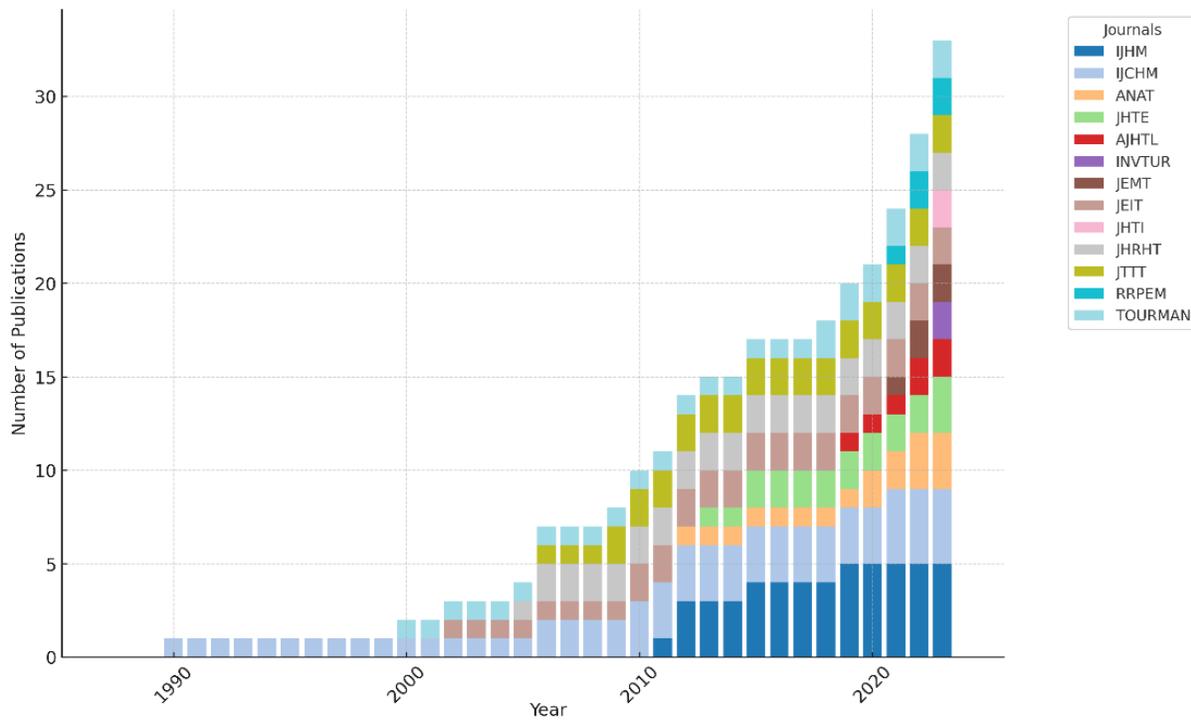

Tourism Management (TOURMAN) stands out significantly in the field of tourism with a H5-Index of 126 and H5-Median of 185. It has the highest CiteScore of 22.9 and is ranked #1/141 in Scopus, alongside a Q1 ranking in WoS and the second-highest JIF of 12.7, indicating its preeminent position in the field. While the International Journal of Hospitality Management (IJHM) is another standout in Tourism with a high H5-Index (130) and H5-Median (194), signaling its strong impact in recent years. It also holds a prominent CiteScore of 18.3, ranking 4th among 141, and has an impressive H-Index (151). Its placement in Q1 across Google Scholar, Scopus, and WoS further underscores its status as a leading publication. The JIF of 11.7 and a #5/58 JIF Rank confirm its influence in the field. Both TOURMAN and IJHM have significant influence and impact within the academic community. IJCHM while not as highly ranked as IJHM or TOURMAN, still maintains a strong presence in the top echelon of research outlets (**Table 3**).

On the other hand, journals like ANAT, JHTE, JHTI, and JTTT hold moderate positions with respect to bibliometric indicators, suggesting their specialized but considerable impact within their respective categories. AJHTL, INVTUR, JEMT, and JHRHT appear to be more niche journals with lower impact scores and quartile rankings, indicating a more focused reach.

**Table 3**

*Core Source Metrics*

| Sources | | Google Scholar | | Scopus | | | | | | | WoS | | | | | | |
|---|---|---|---|---|---|---|---|---|---|---|---|---|---|---|---|---|---|
| Code | Category | H5-Index | H5-Median | Quartile | CiteScore | CiteScore Rank | SJR | SNIP | F-WCI | H-Index | Quartile | Edition | JIF | JIF Rank | JCI | NES | 5-IF |
| 1. IJHM | Tourism | 130 | 194 | Q1 | 18.3 | #4/141 | 2.928 | 2.69 | 2.23 | 151 | Q1 | SSCI | 11.7 | #5/58 | 2.97 | 3.96 | 11.5 |
| 2. IJCHM | Tourism | 92 | 127 | Q1 | 13.6 | #11/141 | 2.5 | 2.07 | 1.8 | 113 | Q1 | SSCI | 11.1 | #6/58 | 2.11 | 2.72 | 9.8 |
| 3. ANAT | Geography | 34 | 49 | Q2 | 4.3 | #177/779 | 0.543 | 0.9 | 1.08 | 38 | - | - | - | - | - | - | - |
| 4. JHTE | Education | 22 | 35 | Q2 | 5.4 | #173/1469 | 0.785 | 1.64 | 1 | 28 | Q1 | ESCI | 2.9 | #123/759 | 1.43 | 0.09 | 2.6 |
| 5. AJHTL | Tourism | - | - | Q4 | 2.1 | #91/141 | 0.216 | 0.38 | 0.28 | 18 | - | - | - | - | - | - | - |
| 6. INVTUR | Tourism | 12 | 16 | Q4 | 1 | #117/141 | 0.19 | 0.37 | 0.25 | 6 | Q4 | ESCI | 0.6 | #120/136 | 0.13 | 0.022 | 0.6 |
| 7. JEMT | Tourism | 30 | 41 | Q4 | 1.9 | #93/141 | 0.199 | 0.5 | 0.31 | 19 | - | - | - | - | - | - | - |
| 8. JEIT | Management | 35 | 60 | Q2 | 4 | #72/214 | 0.621 | 1.35 | 0.98 | 65 | Q3 | ESCI | 2.2 | #278/401 | 0.4 | 0.21 | 2.4 |
| 9. JHTI | Tourism | 31 | 45 | Q2 | 4.5 | #53/141 | 0.641 | 1.09 | 0.94 | 17 | Q2 | ESCI | 3.9 | #44/136 | 0.86 | 0.15 | 3.7 |
| 10. JHRHT | Tourism | 28 | 42 | Q2 | 3.2 | #73/141 | 0.611 | 0.82 | 0.66 | 34 | - | - | - | - | - | - | - |
| 11. JTTT | Education | 20 | 33 | Q2 | 3.6 | #381/1469 | 0.452 | 1.21 | 0.77 | 28 | Q1 | ESCI | 2.5 | #180/759 | 1.19 | 0.07 | 2.3 |
| 12. RRPEM | Education | 28 | 50 | - | - | - | - | - | - | - | Q2 | ESCI | 2 | #255/759 | 0.93 | 0.14 | 1.5 |
| 13. TOURMAN | Tourism | 126 | 185 | Q1 | 22.9 | #1/141 | 3.561 | 3.64 | 3.32 | 236 | Q1 | SSCI | 12.7 | #2/58 | 3.05 | 4.72 | 13.1 |

Note. IJHM: International Journal of Hospitality Management; IJCHM: International Journal of Contemporary Hospitality Management; ANAT: Anatolia; JHTE: Journal of Hospitality & Tourism Education; AJHTL: African Journal of Hospitality, Tourism and Leisure; INVTUR: Investigaciones Turísticas; JEMT: Journal of Environmental Management and Tourism; JEIT: European Journal of Training and Development (formerly Journal of European Industrial Training); JHTI: Journal of Hospitality and Tourism Insights; JHRHT: Journal of Human Resources in Hospitality & Tourism; JTTT: Journal of Teaching in Travel & Tourism; RRPEM: Revista Românească pentru Educație Multidimensională; TOURMAN: Tourism Management

*4.2.2 Key Papers*

**Table 4** presents a selection of the ten most cited papers on the topic of literacies in the tourism labor market, describing the impact and reach of these papers using metrics from Scopus, WoS, and PlumX. The most cited paper, by Chang et al. (2011), explore the impact of human resource management on innovation in hospitality firms in China, analyzing data from 196 hotels and restaurants, suggesting a complex relationship between human resource strategies and innovation outcomes in the hospitality sector, and offering insights into fostering innovation through HR practices in this industry.

**Table 4**

*Core Papers*

| Top Papers | Scopus Metrics | | | | | WoS Metrics | | PlumX Metrics | | |
|---|---|---|---|---|---|---|---|---|---|---|
| | Total Citations (TC) | TC per Year | Normalized TC | F-WCI | Journal Quartile | Total Citations | Journal Quartile | Captures | Social | Views |
| 1 (Chang et al., 2011). | 201 | 15.46 | 2.98 | 4.05 | Q1 | 181 | Q1 | 408 | 18 | - |
| 2. (Asree et al., 2010) | 122 | 8.71 | 1.94 | 4.73 | Q1 | 99 | Q1 | 345 | - | - |
| 3. (Sisson & Adams, 2013) | 118 | 10.73 | 1 | 1.8 | Q1 | - | Q1 | 251 | - | - |
| 4. (Ruhanen, 2006) | 110 | 6.11 | 2.4 | 1.7 | Q2 | - | Q1 | 88 | - | - |
| 5. (H. Kong et al., 2012) | 97 | 8.08 | 1.68 | 4.42 | Q1 | 88 | Q1 | 224 | 18 | - |
| 6. (Tsaur et al., 2010) | 94 | 6.71 | 1.49 | 1.71 | Q1 | 87 | Q1 | 135 | - | - |
| 7. (Daniel et al., 2017) | 89 | 12.71 | 2 | 2.91 | Q1 | 58 | Q1 | 327 | - | - |
| 8. (Zehrer & Mössenlechner, 2009) | 89 | 5.93 | 1 | 2.03 | Q2 | - | Q1 | 118 | - | - |
| 9. (Brophy & Kiely, 2002) | 87 | 3.95 | 1 | 3.02 | Q2 | - | Q3 | 163 | - | 388 |
| 10. (Testa & Sipe, 2012) | 85 | 7.08 | 1.48 | 2.38 | Q1 | 66 | Q1 | 334 | - | - |

The Scopus metrics reveals that this paper has amassed 201 citations since its publication. This paper not only has garnered the most total citations but also maintains a high yearly citation rate (15.46), suggesting its content has remained pertinent over time. The Normalized TC and Field Weighted Citation Impact (FWCI) suggest that the paper's citations are significant when compared to other documents of the same age, subject, and document type. This is a clear leader in terms of citation impact across all metrics, indicating that it has been highly influential in its field.

Asree et al. (2010), with 122 total citations in Scopus and a high normalized TC of 1.94, holds a strong citation record. It has a notable FWCI (4.73), the highest among the list, indicating it is highly cited compared to other papers in the same field. In WoS, it has 99 citations, and in PlumX, it has a high number of captures, suggesting substantial engagement. This paper explores, with data from surveys of 88 hotels in Malaysia, how leadership competency and organizational culture influence their responsiveness to their employees and customers, ultimately affecting revenue.

This study provides original insights by evidencing the impact of leadership and culture within service operations on a hotel's responsiveness and financial success.

The FWCI values across all the papers generally correspond with the total citations and TC per year, but there are exceptions, such as Ruhanen (2006) and Brophy & Kiely (2002), which have higher FWCI values relative to their total citations. This might suggest that while these papers have been cited less overall, their influence is significant when weighted for field-specific factors. In parallel, Daniel et al. (2017) and Tsaur et al. (2010) also stand out for their high TC per year relative to their total citation count, indicating that these papers have quickly become influential in their respective fields despite being more recent than some of the other listed works.

In terms of journal quartile rankings in Scopus, most papers are published in Q1 journals, which denotes a perceived high quality or impact within their subject areas. The WoS metrics largely echo the Scopus metrics, with most papers also published in Q1 journals, but with some lacking total citation data. This lack of data is also explained by differences in database coverage. On the other hand, PlumX captures provide an additional layer of engagement measurement, where Chang et al. (2011) and Testa & Sipe (2012) have particularly high captures, indicating a strong interaction with the document beyond citations, such as readership in Mendeley. Besides, social media metrics (Shares, Likes & Comments in Facebook) and views, as indicated by PlumX for a few papers, such as Chang et al. (2011), Kong et al. (2012), and Brophy & Kiely (2002), suggest wider visibility or public engagement beyond academic citations, which might reflect the broader impact of these papers.

Collectively, these 10 articles are the most influential in the academic study of literacies in the tourism labor market, with robust citation metrics that point to their importance and longevity in the field. They are predominantly published in high-impact Q1 journals and demonstrate considerable engagement from the academic and, in some cases, the wider community.

*4.2.3 Top Authors*

**Table 5** presents a consolidated view of the academic influence and productivity of the most relevant authors, as gauged by some scientometric indicators across Google Scholar, Scopus, and WoS. Tom Baum stands out with the highest number of articles (Baum, 1990; Baum & Devine, 2007; Baum & Thompson, 2007; H.-Y. Kong & Baum, 2006; Solnet et al., 2016), showing not only a significant contribution to the field but also collaborating with other researchers.

**Table 5**

*Core Authors*

| Top Authors | Merged Dataset Metrics | | | Google Scholars metrics | | | |
| --- | --- | --- | --- | --- | --- | --- | --- |
| | Researcher ID | Articles | Timespan | Total Citations | 5-Year Citations | h-index | i10-index |
| Tom Baum | ORCID, SCOPUS, WOS | 5 | 2006-2016 | 19,682 | 8,859 | 73 | 197 |
| Catherine Cheung | ORCID, SCOPUS, WOS | 3 | 2012-2015 | 5,505 | 3,081 | 43 | 79 |
| Clair Haven-Tang | ORCID, SCOPUS, WOS | 2 | 2006-2008 | 1,587 | 756 | 16 | 23 |
| Eleri Jones | SCOPUS, WOS | 2 | 2006-2008 | - | - | - | - |
| Haiyan Kong | ORCID, SCOPUS, WOS | 2 | 2006-2012 | 2,472 | 1,577 | 20 | 28 |
| Lukáš Malec | SCOPUS, WOS | 2 | 2014-2018 | - | - | - | - |
| Inna Rozhi | ORCID, WOS | 2 | 2021-2022 | 70 | 61 | 4 | 2 |

| Top Authors | Scopus Metrics | | | | | WoS Metrics | | |
| --- | --- | --- | --- | --- | --- | --- | --- | --- |
| | Total Citations | Documents | h-index | % First Author | FWCI | Total Citations | Documents | h-index |
| Tom Baum | 6,032 | 203 | 40 | 15% | 4.35 | 3,381 | 121 | 29 |
| Catherine Cheung | 2,463 | 72 | 27 | 29% | 1.45 | 1,964 | 76 | 26 |
| Clair Haven-Tang | 542 | 36 | 11 | 24% | 0.61 | 305 | 24 | 8 |
| Eleri Jones | 2,095 | 60 | 23 | 0% | 0.00 | 947 | 55 | 16 |
| Haiyan Kong | 1,107 | 43 | 19 | 33% | 1.95 | 840 | 36 | 15 |
| Lukáš Malec | 34 | 12 | 4 | 55% | 0.29 | 109 | 20 | 5 |
| Inna Rozhi | - | - | - | - | - | 12 | 6 | 1 |

Baums' five papers investigate the skill requirements and labor dynamics in the hospitality industry across different regions, including transition economies in Asia and established markets like Northern Ireland and China. They examine the impact of globalization on labor markets, the specificity of skills in cultural contexts, and future changes in the sector up to 2030. His research highlights the need for targeted education and training to address skill shortages and to improve service quality in the global hospitality industry. Bam also exhibits a strong citation impact on Google Scholar with 19,682 total citations and an impressive h-index of 73, suggesting high-quality and influential research. Scopus and WoS metrics reinforce this view, showing substantial citations and documents count, with a notable Scopus h-index of 40 and a significant first author percentage, indicating a leading role in research contributions.

Catherine Cheung also displays a notable academic presence with three relevant articles (Kong et al., 2012; Sucher & Cheung, 2015; Yang et al., 2015) and an h-index of 43 on Google Scholar, paired with substantial total citations on both Scopus and WoS. Cheung's first author percentage is reasonable at 29%, reflecting a noticeable role in research authorship, and a fair FWCI score suggests a favorable weighted citation impact. Cheung's papers collectively investigate the development of career management, cross-cultural competency, and employability skills within the hospitality industry. For instance, Kong et al (2012) delves into how hotel career management practices influence employee career satisfaction, mediated by career competency, while Sucher & Cheung (2015) underscore the positive impact of cross-cultural competencies on team performance within multinational hotel companies in Thailand, providing strategic insights for enhancing organizational effectiveness. Lastly, Yang et al (2015) pioneer in empirically establishing a model for employability skills tailored to entry-level hotel staff in China, identifying key skill dimensions essential for job performance. These studies together provide a comprehensive view on the importance of skill development, cultural competence, and career management for sustaining the growth and competitive advantage of the global hospitality sector.

Clair Haven Tang has two relevant articles published jointly with Eleri Jones (Haven-Tang & Jones, 2006, 2008). Clair Haven Tang shows a moderate impact with a Google Scholar h-index of 16 and a Scopus h-index of 11, and the FWCI indicates a modest but noteworthy citation impact within the field. On the other hand, Eleri Jones's metrics reveal a Scopus h-index of 23 out of 60 papers, indicating a focused yet impactful research contribution despite the absence of Google Scholar metrics. Their two papers provide an analysis of the tourism labor market and skills needs in Wales, revealing skills gaps and soft skills deficiencies among job applicants, with employers often resorting to measures such as downsizing or overseas recruitment. They finally suggest that public sector interventions should be more closely aligned with employer demands for skills development to enhance destination competitiveness and service quality.

On the other side, both Lukáš Malec and Inna Rozhi have a more tempered academic impact. However, both Lukáš Malec and Haiyan Kong distinguish themselves by having the highest recorded percentage of first authorship, underlining an important research initiative within this field. In parallel, Kong's scholarly contributions are acknowledged consistently across Scopus and WoS, as evidenced by stable h-index figures.

### 4.3 Co-authorship Network Analysis and Thematic Maps

A co-authorship analysis was carried out using the WoS and Scopus databases to provide evidence on patterns of scientific collaboration to answer the second research question (RQ2).

Several insights can be gleaned by comparing the most representative authors, their occurrence, total link strength, temporal distribution, and average citation metrics. In **Figure 8** the nodes represent individual authors, and the connections (edges) between them indicate collaborative relationships, meaning that the authors have co-authored one or more publications together.

Scopus has a higher maximum number of documents associated with a single author (5) compared to WoS (2), suggesting that some authors in Scopus have a more substantial individual research output. However, the maximum link strength observed in WoS is higher (10) than in Scopus (6), as authorship frequency is greater in WoS, revealing strong collaboration resulting from authors working together on the same publications. The WoS network also exhibits a slightly higher average link strength (2.74) compared to Scopus (2.10). This suggests more frequent or subtly stronger collaborative ties within the WoS network and it further supports the notion of more intense collaborations among certain authors in WoS (refer to the density map in **Figure 8**). Central to the WoS network is Inna Rozhi, who emerges as the figure with the premier total link strength (10), in collaboration with associates including Olena Dutchak, Mykhailo Fomin, Vitalii Honcharuk, Nataliia Horbatiuk, Halyna Humenyuk, Mykola Moskalenko, Iuliia Pologovska, Myhailo Poplavskyi, Yuliia Rybinska, and Valentyna Shchabelska. Conversely, the leading co-authorship cluster in Scopus is apparently orchestrated by Tom Baum, possessing the paramount total link strength (6), and concurrently standing as the most referenced author, hence central to scholarly pursuits.

Within this Scopus network, notable researchers such as Lockstone-Binney (2016), Robinson (2015; 2016) and Solnet (2016) are distinguished for their substantial total link strengths. Aged thirteen years, the average publication year of the main cluster of this network is 2011, while the span of publication years from 2005 to 2016 signals a well-entrenched base indicating a longer-term engagement in the topic, contrasting with the WoS main cluster's relatively transient scope (2021-2022). In general, collaborative research in WoS is more recent on average (2019) compared to Scopus (2017). The data suggest that the WoS network may be capturing newer trends or emerging topics in the field of literacies within the tourism sector.

**Figure 8**

*Co-authorship Network Analysis*

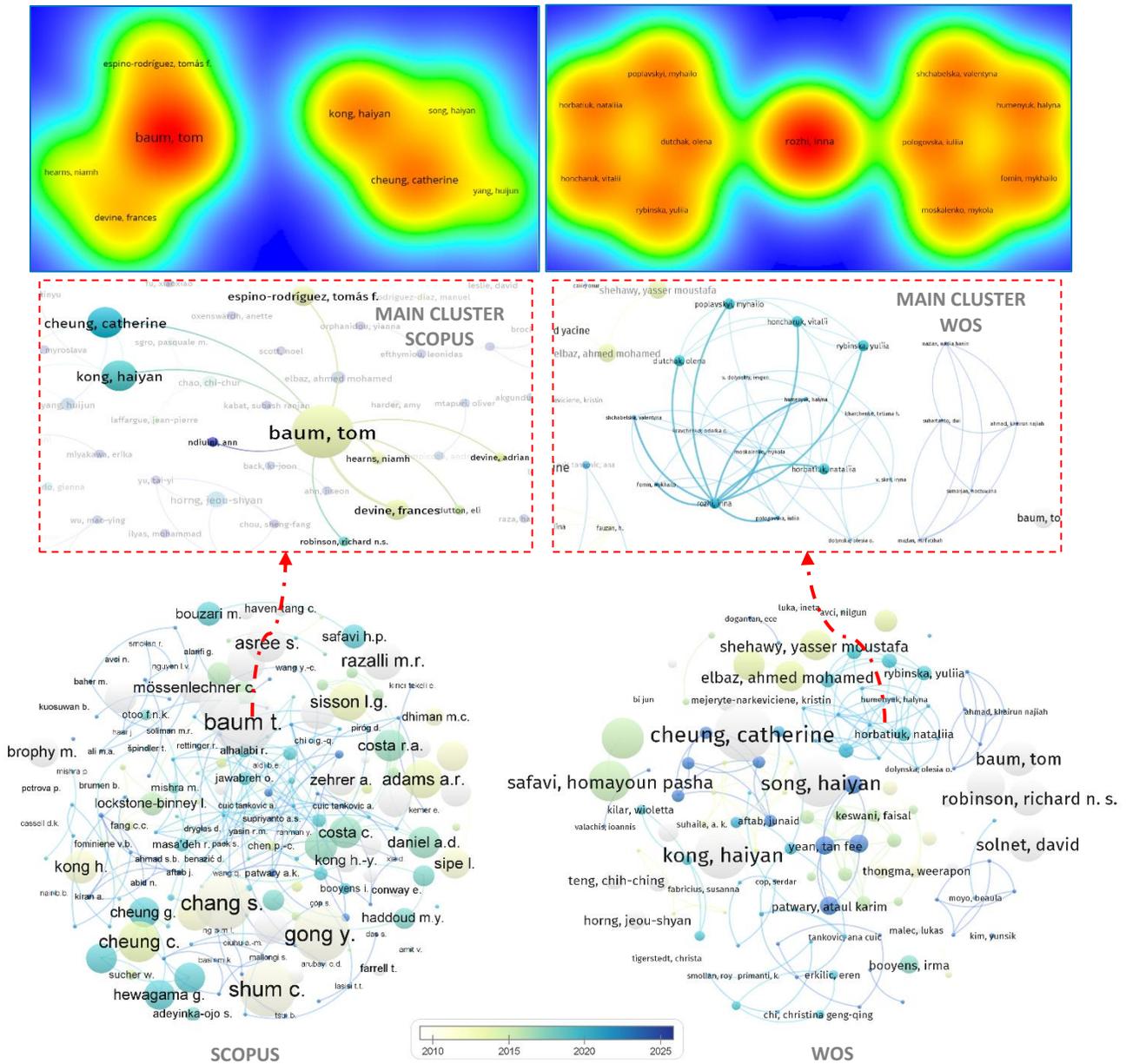

Therefore, to elucidate these possible thematic structures or research topics identified, we employed the use of thematic maps to compare the specific patterns in WoS (**Figure 9**) that stand out in relation to the overall thematic map (**Figure 10**). The overall thematic map (Merged Dataset) emphasizes well-established educational and competency themes such as "higher education",

"curriculum design", "employee performance", "employee competencies" or "transformational leadership", showing these topics as highly dense or central to current discussions. In contrast, the WoS map reveals an inclination towards "serious games", "service learning", "sustainable tourism" and "emotional intelligence". These topics, beyond their significance in the WoS network, also share multiple elements that enable us to classify them as critical educational methodologies within the context of tourism literacies. For example, serious games use game design principles to facilitate learning (Lalicic & Weber-Sabil, 2022), whereas service learning integrates learning with community service, fostering a deeper understanding of academic concepts and their practical applications (Jamal et al., 2011). Furthermore, sustainable tourism uses experiential learning to promote cultural awareness, environmental stewardship, and community development while emotional intelligence helps individuals develop essential life skills, such as self-awareness, empathy, and effective communication (Ornstein & Nelson, 2006; Tekeli & Ozkoc, 2022)

Nevertheless, despite the conceptual intersections among these topics, cross-collaboration among authors within the co-authorship network remains limited. For instance, the application of serious games to sustainable tourism literacy has been minimally explored, even with its advantages as a groundbreaking method. This could be attributed to the fact that serious gaming is an interdisciplinary co-creative methodology, necessitating the collaboration of researchers from diverse fields of knowledge as stated by Lalicic & Weber-Sabil (2022). Hence, the evolution of the topics also indicates that collaboration among authors could benefit from a multidisciplinary approach that integrates primarily educational, psychological, computer science, and environmental sustainability perspectives.

Beyond the above key themes, **Figures 9** and **10** show that the main "Emerging Themes" are "soft skills", "communication skills" and "intercultural competence", while "managerial competencies" is the main "Declining Theme". In the context of hospitality and tourism, soft skills such as empathy, teamwork, flexibility, and problem-solving are crucial, as they enable employees to deliver superior customer service, adapt to various situations, and work effectively in diverse teams (Moura et al., 2020; Tankovic, Vitezic, et al., 2023; Vasconcelos et al., 2022).

**Figure 9**

*WoS Thematic Map*

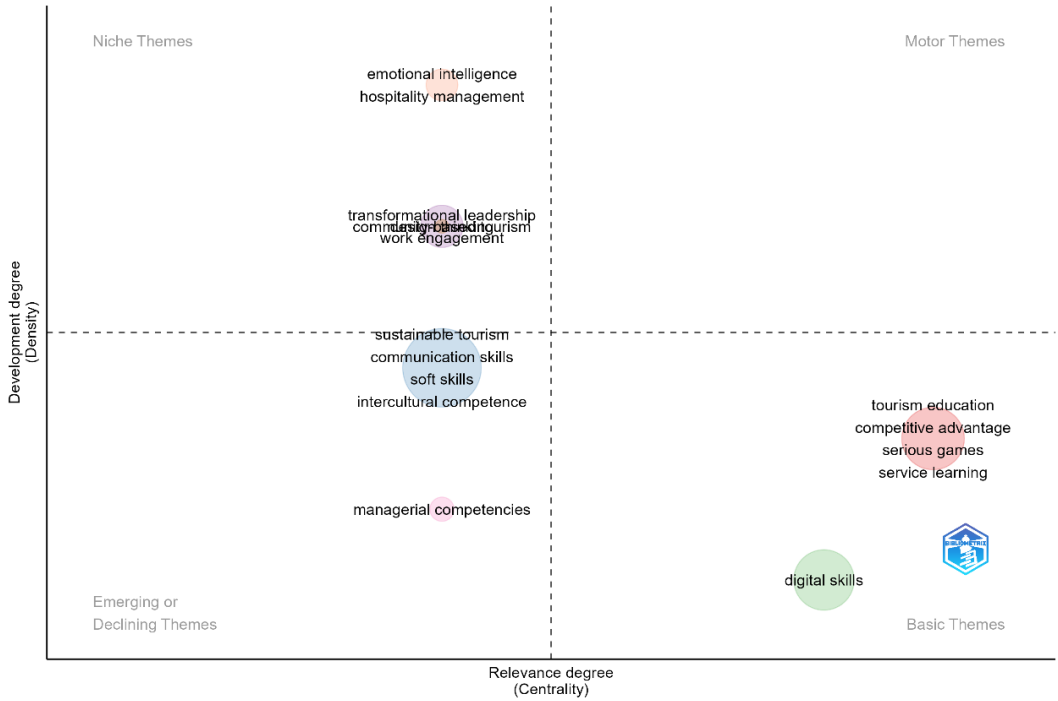

**Figure 10**

*Overall Thematic Map*

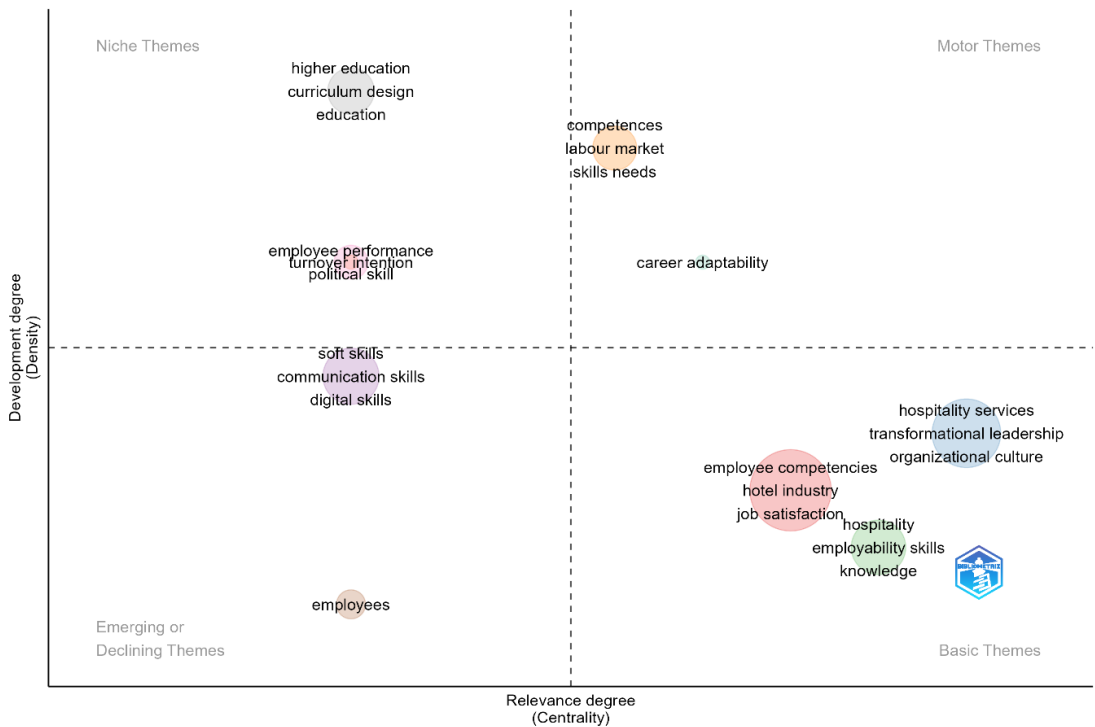

Similarly, good communication skills improve interactions with customers, enhance guest experiences, and facilitate the smooth operation of tourism services (Tankovic, Kapes, et al., 2023; Tankovic, Vitezic, et al., 2023), whereas intercultural competence involves understanding cultural differences, showing respect for diverse traditions and customs, and being able to engage effectively with people from diverse cultural backgrounds (Apelt et al., 2023; Fan et al., 2022), which not only helps to provide personalized service to a global clientele, but also to navigate and manage a culturally diverse workforce.

In addition, an important difference between both thematic maps should be noted. In WoS, digital skills were positioned at the edge of the "Basic Themes" quadrant, suggesting that it is a cross-cutting and central theme within the network, although it is still underdeveloped. Conversely, in the overall patterns (Merged Dataset), digital skills were squarely in the "Emerging Themes" quadrant, indicating an evolving area of interest that has not yet reached prominence. This suggests that research on digital skills is still in a transitional phase within tourism. With the recent expansion of new technologies, robotics, internet of things and artificial intelligence, the insertion of tourism in the context of digital skills will become increasingly important.

All these findings share commonalities with recent research conducted by Au-Yong-Oliveira et al (2024). They identified "green skills", "digital skills", "soft skills", "emotional intelligence" and "self-motivation" as key competencies in today's tourism industry. Their study focused on a bibliometric analysis and literature review of these competencies specifically within an educational context, whereas our focus has been on the broader context of the tourism labor market. By contrasting their findings with ours, three key literacies with implications not only for academia but also for the current tourism labor market can be identified: "Digital Literacy", required for effective use of digital technology, software tools, social networks, etc.; "Social & Emotional Literacy", encompassing soft skills and emotional intelligence; and "Environmental Literacy" directly tied to green skills, which includes knowledge of climate change, sustainability, and regenerative tourism.

In a way, these three literacies form the main trending pattern in current research in this area and, therefore, constitute a reference for training programs, skills building, curriculum optimization, as well as for reducing entry barriers to the tourism labor market. Similarly, these results confirm the findings of Au-Yong-Oliveira et al (2024), outlining the need to guide future studies to deepen these literacies in a comprehensive way, promoting greater collaboration and interdisciplinary integration, since the point of simultaneous convergence of these three literacies, as well as their evolution and implications for the future of tourism, has not yet been clearly explored by previous studies.

## 5. Conclusion

Key findings from this study include the observation of a moderate increase in research pertaining to literacies within tourism, evident across both Scopus and WoS databases. Even though four possible categories were identified in which core sources are indexed (Tourism, Education, Geography, and Management), the analysis of performance metrics indicated that those journals which exhibit both higher scientific impact and production levels are concentrated in the "Tourism" category. Particularly, Tourism Management (TOURMAN) and International Journal of Hospitality Management (IJHM) are the leading core journals, while IJHM and International Journal of Contemporary Hospitality Management (IJCHM) stand out as the top publishers in this field. These data suggest that the most suitable and prestigious journal for publishing in this specific topic could be IJHM (JCR Q1). However, if the journal's prestige is not such a significant factor due to the nature of the research, Journal of Hospitality and Tourism Insights (JHTI) may be considered for publication in English, and Investigaciones Turísticas (INVTUR) if publishing in Spanish, as these journals have demonstrated the most rapid growth over the past 3 years in this research domain.

Chang et al. (2011), Asree et al. (2010), and Sisson & Adams (2013) are the most cited papers. These studies emphasize the importance of human resources management practices and organizational culture in driving innovation, responsiveness, and performance within the hospitality industry, highlighting the significance of selecting and training multi-skilled employees, leadership competency, and fostering a supportive organizational culture as key factors for success. Each study makes use of empirical data and rigorous methodologies (e.g., structural equation modeling) to investigate their respective hypotheses, contributing original insights into the field of hospitality management.

On the other hand, by establishing a link between the results of the core authors' metrics analysis and co-authorship analysis, it was possible to verify that the most influential author in this field of study is Tom Baum. Baum not only has the highest number of total articles as an author or co-author but also exhibits a strong citation impact and h-index. This author also plays a significant role in the co-authorship network, occupying the central position in the main Scopus cluster with a higher Total link strength, also having a considerable impact within the WoS network, and demonstrating extraordinarily high performance in Google Scholar metrics. Overall, Scopus outperforms in terms of the maximum quantity of publications attributed to a single author and secures robust citation rates. In contrast, the WoS framework shows a subtle advantage in the average strength of connections, suggesting a slightly more profound collaboration intensity among its authors. This indicates a nuanced increase in collaborative dynamics within the WoS database.

Our findings suggest that the WoS network may be capturing newer trends or emerging topics. Three key literacies with implications not only for academia but also for the current tourism labor market were identified: "Digital Literacy", "Social & Emotional Literacy", and "Environmental Literacy". Mastery of these literacies enables staff to respond more effectively to the diverse needs and expectations of international tourists, leading to higher satisfaction rates and repeat business. However, despite the growing body of existing research on these key literacies, cross-collaboration between authors within the co-authorship network remains limited and future studies are needed to have greater impact in analyzing how the conceptual intersections of these 3 literacies are shaped within the tourism labor market and what their implications are for future job demand in the sector.

Essentially, this book chapter addressed literacies within tourism from a bibliometric standpoint. The findings and their discussion can serve as a reference for researchers planning future research on this topic. It can be particularly useful as a pointer to in-depth exploration of core papers in literature reviews using both WoS and Scopus. Likewise, the implications provided for potential journals could be valuable to researchers when considering where to publish their work in this field.

This book chapter also holds practical implications for the tourism sector. It offers a focused selection of key scientific contributions, supporting tourism managers in identifying research that could improve decision-making, workforce assessments in the digital era, and the development of targeted skill-building educational strategies. Although this book chapter has an exploratory focus, to the best of our knowledge this is one of the first bibliometric studies on literacies in the context of the tourism labor market.

**5.1 Study limitations and guidelines for future research**

Based on our methodology and results, several limitations and areas for future research emerge. Firstly, the moderate overlap between Scopus and WoS databases suggests a scattering of research that derives in the need to expand the database scope to uncover additional relevant literature. Therefore, expanding the search to include additional resources indexed in Google Scholar, Dimensions, Microsoft Academic or Semantic Scholar, for instance, could help in identifying more relevant studies that were not captured due to the limitations of Scopus and WoS. Also, this study verified that the overlap rate between the databases stands at 35.71%. This figure is neither so low as to indicate complete independence between the databases nor so high as to suggest almost identical coverage. Nonetheless, given that the analysis of differences between the two databases included performance metrics and collaboration network analyses, it is essential for future studies to deeply investigate the reasons behind the distinct differences beyond mere coverage disparities, thereby better identifying the research most suited to each database.

Secondly, the reliance on citation metrics, altmetrics or co-autorship network analysis as indicators of influence may not fully capture the nuanced impact of research within the field, suggesting that future studies should consider alternative approaches such as detailed accounts of how research has been applied in real-world scenarios, demonstrating tangible outcomes and societal benefits, gathering insights from field experts on the perceived influence and utility of research, offering a peer perspective on impact, and evaluating the contribution of research to advancing knowledge, policy, and practice through systematic reviews or meta-analyses.

Given that research collaborations in WoS are characteristically more recent and simultaneously more intense by having a higher average link strength, it would be interesting for future research directions to explore possible points of change in the theory. Therefore, it will be interesting to study trending topics to gain new perspectives not only at present but also considering its future evolution. For example, to know to what extent Digital Literacies will impact the tourism sector in the future or to know which new Literacies may emerge in the coming years.

Finally, the co-authorship network analysis revealed a relatively narrow collaboration network in general, suggesting that future efforts could focus on fostering broader interdisciplinary collaborations to enrich the field's diversity of thought and methodology.

### 5.2 Funding

This research received support from the Agència de Gestió d'Ajuts Universitaris i de Recerca (AGAUR) and was co-funded by the European Union, under Grant 2023 FI-1 00622, and takes part of the Grants 2023PFR-URV-114, 2022PFR-URV-4 and Grant PID2021-122575NB-I00 funded by MCIN/AEI/ 10.13039/501100011033/ by "ERDF A way of making Europe".

*Notes in Artificial Intelligence and Lecture Notes in Bioinformatics)*, *14111 LNCS*, 470 – 488. https://doi.org/10.1007/978-3-031-37126-4_31

Higaki, A., Uetani, T., Ikeda, S., & Yamaguchi, O. (2020). Co-authorship network analysis in cardiovascular research utilizing machine learning (2009–2019). *International Journal of Medical Informatics*, *143*. https://doi.org/10.1016/j.ijmedinf.2020.104274

Idris, I., Suyuti, A., Supriyanto, A. S., & As, N. (2022). Transformational leadership, political skill, organizational culture, and employee performance: A case from tourism company in Indonesia. *Geojournal of Tourism and Geosites*, *40*(1), 104 – 110. https://doi.org/10.30892/GTG.40112-808

Indahyani, T., Dewanti, N. R., & Meliana, S. (2023). Implementation of digital transformation in design education learning to support the sustainability of tourism area. *2023 IEEE 9th International Conference on Computing, Engineering and Design, ICCED 2023*. https://doi.org/10.1109/ICCED60214.2023.10425742

Jamal, T., Taillon, J., & Dredge, D. (2011). Sustainable tourism pedagogy and academic-community collaboration: A progressive service-learning approach. *Tourism and Hospitality Research*, *11*(2), 133–147. https://doi.org/10.1057/thr.2011.3

Jeong, J. Y., Karimov, M., Sobirov, Y., Saidmamatov, O., & Marty, P. (2023). Evaluating culturalization strategies for sustainable tourism development in Uzbekistan. *Sustainability (Switzerland)*, *15*(9). https://doi.org/10.3390/su15097727

Ji, W., Yu, S., Shen, Z., Wang, M., Cheng, G., Yang, T., & Yuan, Q. (2023). Knowledge mapping with Citespace, Vosviewer, and Scimat on intelligent connected vehicles: Road safety issue. In *Sustainability (Switzerland)* (Vol. 15, Issue 15). https://doi.org/10.3390/su151512003

Johnson, P. C. (2014). Cultural literacy, cosmopolitanism and tourism research. *Annals of Tourism Research*, *44*, 255–269. https://doi.org/10.1016/j.annals.2013.10.006

Kim, Y. (2023). Examining the impact of frontline service robots service competence on hotel frontline employees from a collaboration perspective. *Sustainability*, *15*(9). https://doi.org/10.3390/su15097563

Koç, H., Karacabey, F. A., & Demir Yurtseven, E. (2023). Developing a scale for tourism literacy: validity and reliability study. *Current Issues in Tourism*. https://doi.org/10.1080/13683500.2023.2267731
37